\documentclass[review,times,onecolumn,round,authoryear,square]{elsarticle}

\usepackage{amssymb,amsmath}
\usepackage{graphics,graphicx}
\usepackage{url,theorem}
\usepackage{caption}
\usepackage{xcolor}
\usepackage[version=3]{mhchem}

\journal{Chemical Engineering Science}

\newcommand{\torol}[1]{}
\newcommand{\emp}[1]{\textbf{#1}}
\newcommand{\mma}[1]{{\bf\tt #1}}

\newcommand{\cb}{\ifmmode \mathbf{c}\else \textbf{c}\fi}

\newcommand{\R}{\mathbb{R}}

\newcommand{\ReactionKinetics}{\ifmmode {\textbf{\texttt{ReactionKinetics}}}
\else {\bf\tt ReactionKinetics}\fi}

\newcommand{\FHJg}{Feinberg--Horn--Jackson graph}

\newcommand{\Mma}{\textit{Mathematica}}

\newcommand{\N}{\mathbb{N}}
\theorembodyfont{\upshape}

\newtheorem{Exa}{Example}

\begin{document}
\begin{frontmatter}
\title{Structural Analysis of Combustion Mechanisms}
\author[bmeanal,elte]{J. T\'oth\corref{cor1}\fnref{fn1,fn2,fn3}}
\ead{jtoth@math.bme.hu}
\author[bmestoch]{A. L. Nagy\fnref{fn1,fn3,fn4}}
\ead{nagyal@math.bme.hu}
\author[elte]{I. Gy. Zs\'ely\fnref{fn1,fn3}}
\ead{zselyigy@gmail.com}

\address[bmeanal]{Department of Analysis,
Budapest University of Technology and Economics,
Egry J. u. 1., Budapest, Hungary, H-1111}

\address[bmestoch]{Department of Stochastics,
Budapest University of Technology and Economics,
Egry J. u. 1., Budapest, Hungary, H-1111}

\address[elte]{Laboratory for Chemical Kinetics of the
Department of Physical Chemistry,
E\"otv\"os Lor\'and University,
P\'azm\'any P. s\'et\'any 1/A.,
Budapest, Hungary, H-1117}


\fntext[fn1]{Partially supported by the Hungarian National Scientific Foundation,
No. 84054 and No. 84060.}
\fntext[fn2]{This work is connected to the scientific program of the "Development of quality-oriented and harmonized R+D+I strategy and functional model at BME" project.
This project is supported by the New Sz\'echenyi Plan (Project ID: T\'AMOP-4.2.1/B-09/1/KMR-2010-0002).}
\fntext[fn3]{Partially supported by the COST Action CM901: Detailed Chemical Kinetic Models for Cleaner Combustion.}
\fntext[fn4]{The results discussed below are supported by the grant T\'AMOP-4.2.2.B-10/1--2010-0009.}
\cortext[cor1]{Corresponding author}
\begin{abstract}
$39$ detailed mechanisms for combustion of hydrogen, carbon monoxide and methanol are investigated
using \mma{ReactionKinetics}, a {\it Mathematica} based package.
The obtained results in most cases do not depend on the choice of reaction rate coefficients,
the methods only use the underlying sets of reaction steps,
thus the results are robust and general in a certain sense.
These investigations can be used before or in parallel with usual numerical investigations,
such as pathway analysis, sensitivity analysis, parameter estimation or simulation.

The considered hydrogen mechanisms shared 90\% of common reaction steps.
The CO combustion mechanisms show a larger variety both in species and in reaction steps.
There exist only a few methanol combustion mechanisms;
the big differences between them shows that the modelling community
is only at the very beginning of exploring this process.

The package and the methods may be useful for automatic mechanism generations, testing, comparing and reduction of mechanisms as well, especially in the case of large systems.
\end{abstract}


\end{frontmatter}
\section{Introduction}
The reaction steps of hydrogen and carbon monoxide combustion form a
central part of the high temperature combustion of all hydrocarbons
and oxygenates, see e.g. \cite{reed}.
Also, hydrogen is an important fuel itself for different applications (e.g. rocket propulsion)
and in the context of a carbon-free economy as well as of safety issues.
In the recent years, there has been an increased interest in studying the
combustion of fuel mixtures consisting of carbon monoxide and
hydrogen, referred to as "wet CO" or syngas. These fuels can be
produced from coal and biomass via gasification, and are considered to
be a promising option towards cleaner combustion technologies for
power generation. Oxygenated organic compounds have been proposed as
alternative fuels in order to improve the fuel properties and reduce
particulates and \ce{NO_x} emissions.
Methanol, the most simple alcohol, is one of the most important oxygenated additives
due to its high oxygen content and the lack of C--C bonds.

In the present paper we investigate the combustion of hydrogen, carbon monoxide and methanol:
three phenomena important both from theoretical and practical points of view in combustion.

The approach we use is absolutely structural in the sense that none of the results
depend on the values of the rate coefficients (cf. \cite{beck}).
We might say that we are going to discover possibilities instead of quantitatively dealing with individual mechanisms.
To put it another way, we are going to raise questions to be answered by the chemist,
rather than answering them.

In a previous paper we presented a \textit{Mathematica} based program package called \ReactionKinetics\
\citep{nagypapptoth}  aimed at symbolic and numerical treatment of chemical mechanisms.
The package is especially useful when the numbers of species and reaction steps are
larger than to allow manual investigations, i.e. if one has dozens or even thousands
of species and reaction steps.
After the publication of the previous version we made the package capable of reading
CHEMKIN files by \mma{CHEMKINImport}, added dozens of new functions
such as e.g.
\begin{itemize}
\item\mma{CHEMKINExport},
\item\mma{MaxFHJWeaklyConnectedComponents},
\item\mma{MinFHJWeaklyConnectedComponents},
\item\mma{MaxFHJStronglyConnectedComponents},
\item\mma{MinFHJStronglyConnectedComponents},
\item\mma{FilterReactions},
\end{itemize}
and made the package compatible with Version $9$ of \textit{Mathematica}.

The structure of our paper is as follows. 
In Section 2 the mechanisms to be investigated are described. 
Results are shown in Section 3.
The necessary mathematical background is relegated to the Appendix.

Finally, two electronic supplements are added.
Firstly, a \textit{Mathematica} notebook  showing all the details
of the calculations  which may be really useful for those interested
in combustion modeling but of minor interest for the general audience.
Some of the resulting figures are also given there.
The calculations can be reproduced by using the package itself.
It can be downloaded from the following page:
\begin{center}
\url{http://www.math.bme.hu/~jtoth/CES2013}.
\end{center}
The data can either be collected from the original authors, or
from our database to be built in the near future.
Secondly, we also attach the (very long) PDF manuscript
of our notebook which allows to passively follow what we have done,
but this version does not need the \Mma\ program.

\section{Selected mechanisms of combustion of hydrogen, carbon monoxide and methanol}
The simplest chain branching combustion reaction, the oxidation of hydrogen is already
a much more complex system than the Mole and Robertson reactions discussed in the Appendix.
It is a common misconception that the chemistry of these low-order systems is well understood.
However, \cite{zselyolmpalvolgyivarganagyturanyi} showed recently in
a comprehensive mechanism comparison paper
that this is not the case. The description of the experimental data is still
not satisfactory and some of the recently published reaction mechanisms perform
worse than older ones.
Similar comparison was done by \cite{olmzselyvarganagyturanyi}
for the oxidation of carbon monoxide.
In this work we utilize the mechanism collection of these papers,
but focus on the structural differences of the mechanisms.
By extending the investigations to some detailed methanol mechanisms we show
that the suggested formal mathematical handling is still applicable
for even larger kinetic systems.
The phenomena are more and more complex as we proceed from
hydrogen through carbon monoxide to methanol.
Correspondingly, the mechanisms are larger and more and more diverse.

\section{On the structure of the selected combustion mechanisms}
Even the simplest mechanisms for combustion usually contain dozens of species and of reaction steps, therefore
we can only show selected parts of the results here, e.g.
Volpert graphs of the investigated reactions are not shown here as the figures themselves are not useful,
they can only be used for calculating the Volpert indices.
However, we have shown the Volpert graphs together with Volpert indices for two simple reactions
in the Appendix.
\subsection{Hydrogen}
As a starting point the basic data of the investigated mechanisms are presented in Tables \ref{tab:hbasic1} and \ref{tab:hbasic2}.

\begin{center}
\captionof{table}{Basic data of the investigated hydrogen combustion mechanisms I.}
\label{tab:hbasic1}
\begin{tabular}{|l|l|l|l|l|}
\hline
{\bf Mechanism}&{\bf Reference}&$M$&$R$&$\delta=N-L-S$\\
\hline
Ahmed2007&\cite{ahmedmaussmoreaczeuch}&
8&38&$29-11-6=12$\\
\hline
Burke2012&\cite{burkechaosjudryerklippenstein}&
8&38&$31-12-6=13$\\
\hline
CRECK2012&\cite{healykalitanaulpetersenbourquecurran}&
8&37&$29-11-6=12$\\
\hline
Dagaut2003&\cite{dagautlecomtemieritzglarborg}&
8&42&$31-12-6=13$\\
\hline
Davis2005&\cite{davisjoshiwangegolfopoulos}&
8&40&$31-12-6=13$\\
\hline
GRI30&\cite{smithgoldenfrenklachmoriaryeiteneergoldenbergbowmanhansonsonggardinerlissianski}&
8&40&$31-12-6=13$\\
\hline
Hong2011&\cite{hongdavidsonhanson}&
8&40&$31-12-6=13$\\
\hline
Keromnes2013&\cite{keromnesmetcalfeheuferdonohoedassungherzlernaumanngriebelmathieukrejcipetersenpitzcurran}&
9&42&$32-12-7=13$\\
\hline
Konnov2008&\cite{konnov}&
8&42&$31-12-6=13$\\
\hline
Li2007&\cite{lizhaokazakovchaosdryerscire}&
8&38&$31-12-6=13$\\
\hline
\end{tabular}
\end{center}
\begin{center}
\captionof{table}{Basic data of the investigated hydrogen combustion mechanisms II.}
\label{tab:hbasic2}
\begin{tabular}{|l|l|l|l|l|}
\hline
{\bf Mechanism}&{\bf Reference}&$M$&$R$&$\delta=N-L-S$\\
\hline
NUIG2010&\cite{healykalitanaulpetersenbourquecurran}&
8&38&$31-12-6=13$\\
\hline
OConaire2004&\cite{oconairecurransimmiepitzwestbrook}&
8&38&$31-12-6=13$\\
\hline
Rasmussen2008&\cite{rasmussenhansenmarshallglarborg}&
8&40&$31-12-6=13$\\
\hline
SanDiego2011&\cite{sandiego}&
8&42&$31-12-6=13$\\
\hline
SaxenaWilliams2006&\cite{saxenawilliams}&
8&42&$31-12-6=13$\\
\hline
Starik2009&\cite{stariktitovasharipovkozlov}&
9&52&$41-16-7=18$\\
\hline
Sun2007&\cite{sunyangjomaaslaw}&
8&40&$31-12-6=13$\\
\hline
USC2007&\cite{wangyoujoshidavislaskinegolfopouloslaw}&
8&40&$31-12-6=13$\\
\hline
Zsely2005&\cite{zselyzadorturanyi}&
8&42&$31-12-6=13$\\
\hline
\end{tabular}
\end{center}
\begin{description}
\item[Number of species, $M$]
All mechanisms contains the same (core) set of species:
\ce{H}, \ce{H2}, \ce{H2O}, \ce{OH}, \ce{H2O2}, \ce{HO2}, \ce{O2}, \ce{O}.
The Keromnes2013 mechanism, formally, contains $h\nu$ as a species,
but this is only a description for the photoexcitation in a photochemical reaction step.

This mechanism is the only one which contains the excited \ce{OH} species (\ce{OHEX}) to describe
some ignition delay experiments better.

There is another mechanism (Starik2009) which contains an additional species, ozone.
It is quite unique to include this species in a reaction mechanism intended
to be used for the description of combustion processes.

\torol{
******************
Before turning to the mechanisms relevant to the viewpoint of kinetics we discard the
third bodies present in the data files. It may not be so important, ISTVAN?? but it
is interesting that the only third body remaining after this cleaning
as an internal species taking
part in some reversible reaction steps as shown in Table \ref{tab:hthirdbody}
below  is \ce{N2}.
\begin{center}
\captionof{table}{Reaction steps containing the third body \ce{N2}
as internal species}
\label{tab:hthirdbody}
\begin{tabular}{|l|l|}
\hline
Reaction step&Mechanism\\
\hline
\ce{H + N2 + O2 <=> HO2 + N2}&GRI30\\
\hline
\ce{H2 + N2 <=> 2H + N2}&Hong2011, Konnov2008,\\
&Rasmussen2008, Sun2007\\
\hline
\ce{N2 + OHEX <=> N2 + OH}&Keromnes2013\\
\hline
\end{tabular}
\end{center}
*****************
}

\item[Number of reaction steps, $R$]
The number of reaction steps, $R$, varies between 37 and 44, except Starik2009,
where this number is 52. Thus---as C. K. Law reported
in his comprehensive review paper
\citep{law}---the number of reactions is approximately 5 times larger than
the number of species (except the mentioned case).

\item[Deficiency, $\delta$]
The number of complexes, $N$, varies between 29 and 32, except Starik2009, where this number is 41.
The number of weakly connected components is either 11 or 12,
except again Starik2009, where this number is 16.
The preliminary data suggest that Starik2009 is structurally richer than the other mechanisms.

The deficiencies are large, neither the zero deficiency theorem,
nor the one deficiency theory can be applied.
\item[Weak reversibility and acyclicity]
None of the reactions have an acyclic Volpert graph,
as all the reactions, except in CRECK2012, are fully reversible.
Accordingly, all the reactions are weakly reversible, except again CRECK2012,
which has a single \emp{irreversible step}:
\ce{H2O2 + O -> HO2 + OH}.
\end{description}

\subsubsection{Representations of mechanism classes}
\begin{figure}[!hb]
\begin{center}
\includegraphics[width=0.55\paperwidth]{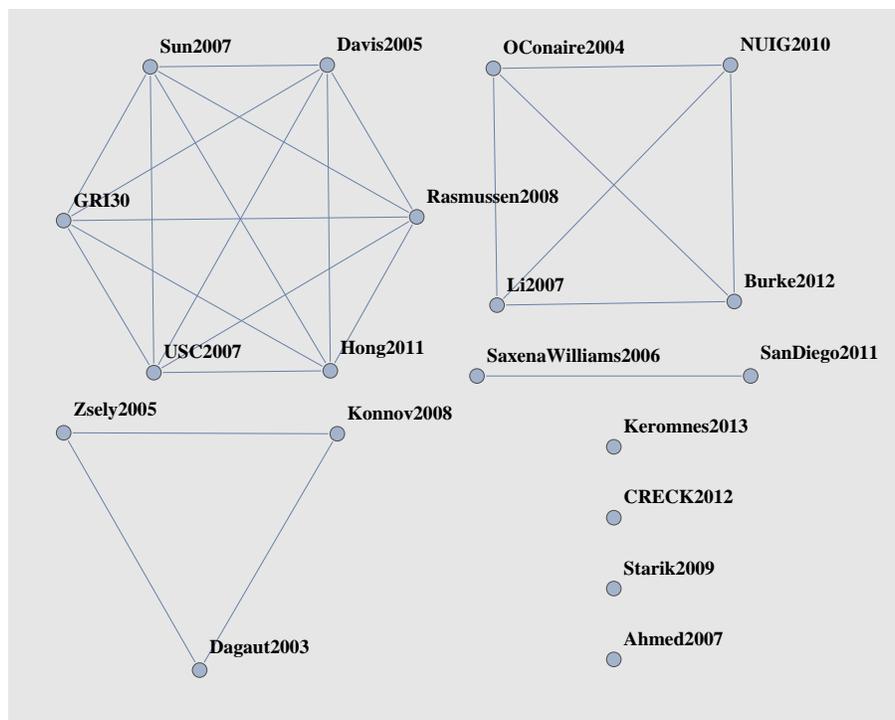}
\caption{Classes of hydrogen combustion mechanisms.
Mechanisms with the same structure are connected with an edge of the graph}
\label{fig:modelclasses}
\end{center}
\end{figure}

From our---let us emphasize: structural---point of view not all the mechanisms in
Tables \ref{tab:hbasic1} \& \ref{tab:hbasic2} are different,
one has classes with exactly the same structure if
the values of reaction rate coefficients are disregarded.
(To put it another way: the underlying complex chemical reaction is the same.)
The classes are shown in Fig. \ref{fig:modelclasses}.

During mechanism development one of the first steps is the decision of which species
should be included in the mechanism.
After this, the reaction steps and the best possible (to the best knowledge of the authors)
set of parameters are selected.
This last step, the assignment of the rate parameters forms the largest part
of a mechanism development work.
However, we have to keep it in mind that the parameter set
corresponds to the previously fixed structure of the model.
Therefore, it is important to compare the mechanisms from the point of view
of their structures. Figure \ref{fig:modelclasses} is a demonstrative example that the
currently published hydrogen combustion mechanisms are different
not at the level of the parameters, but already in their general structures
(i.e. the reaction steps underlying in the mechanism).
It is interesting to see that a significant number of reaction mechanisms
in this collection kept the structure of the old GRI1999.
It is also interesting, that when the mechanisms are updated
most of the authors do not modify their structures
(see the reaction mechanisms coming from the same research group,
e.g. SaxenaWilliams2006 and SanDiego2011, or Li2007 and Burke2012,
or OConaire2004 and NUIG2010). Clinging to potentially outdated structures can be
one of the possible pitfalls in mechanism development.
The structural relationship the OConaire2004, Li2007, NUIG2010 and Burke2012 is
obvious, as they are based on some older reaction mechanisms of Dryer's group
\citep{muellerkimyetterdryer}.

However, this finding allows us to choose a single mechanism from the classes, and we remind the reader that from now on
Davis2005 also represents GRI30, Hong2011, Rasmussen2008, Sun2007, USC2007;
Burke2012 also represents Li2007, NUIG2010, OConaire2004;
and Dagaut2003 also represents Konnov2008 and Zsely2005.\\
SanDiego2011 also represents SaxenaWilliams2006; whereas each of Ahmed2007, CRECK2012, Keromnes2013 and Starik2009
form a separate class. The representatives have been selected by the caprice of the alphabet.

Given that the initial species in the case of hydrogen combustion mechanisms are \ce{H2} and \ce{O2},
the Volpert indices of the species are displayed in Tables \ref{tab:ahmedvolpert}, \ref{tab:burkevolpert}, \ref{tab:creckvolpert}, \ref{tab:dagautvolpert}, \ref{tab:keromnesvolpert} and \ref{tab:starikvolpert}.
The Volpert indices of the reaction steps will only be given in the case of Burke2012 (Table \ref{tab:burkevolpertrsteps}).

\subsubsection{Ahmed2007}
\begin{figure}[!ht]
\begin{center}
\includegraphics[width=0.45\paperwidth]{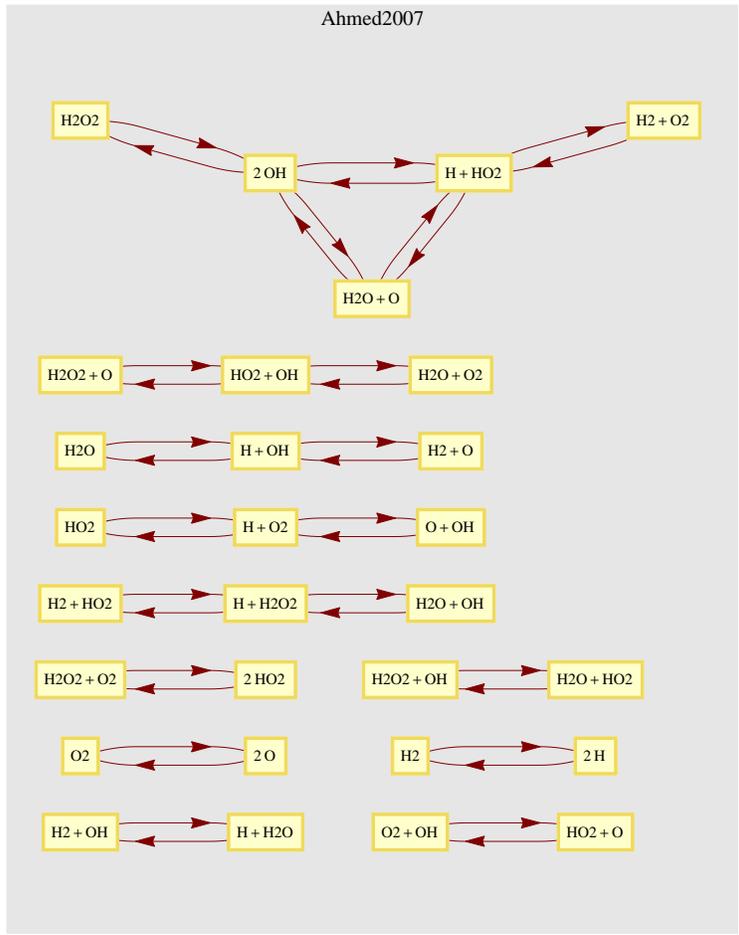}
\caption{The Feinberg--Horn--Jackson graph of Ahmed2007}
\label{fig:AhmedFHJ}
\end{center}
\end{figure}

As an illustration, the Feinberg--Horn--Jackson graph of the Ahmed2007 mechanism is shown in Fig. \ref{fig:AhmedFHJ}.

One can also investigate the maximal (weakly) connected components of the Feinberg--Horn--Jackson graph.
It turned out that this substructure is rather stable: the maximal (weakly) connected components of
Ahmed2007-, Davis2005-, SanDiego2011- and Starik2009-type mechanisms are the same.
Also, in the case of Burke2012-, CRECK2012- and Keromnes2013-type mechanisms we get the same component.
And finally, Dagaut2003-type mechanism is special, its \FHJg\ is a kind of enlargement of the previous graphs.
Chemically, \ce{OH} has four different channels to be transformed
as opposed to three in the other mechanisms. Let us mention that the three most important radicals in combustion: \ce{H}, \ce{O} and \ce{OH} form a full triangle in most of the maximal connected components except the second series:
Burke2012-, CRECK2012- and Keromnes2013-type mechanisms.
\begin{center}
\captionof{table}{Volpert indices of the species in Ahmed2007}
\label{tab:ahmedvolpert}
\begin{tabular}{|l|l|}
\hline
index&species\\
\hline
0&\ce{H2}, \ce{O2}\\
1&\ce{H}, \ce{HO2}, \ce{O}\\
2&\ce{H2O2}, \ce{H2O}, \ce{OH}\\
\hline
\end{tabular}
\end{center}

\subsubsection{Burke2012}
\begin{center}
\captionof{table}{Volpert indices of the species in Burke2012}
\label{tab:burkevolpert}
\begin{tabular}{|l|l|}
\hline
index&species\\
\hline
0&\ce{H2}, \ce{O2}\\
1&\ce{H}, \ce{HO2}, \ce{O}\\
2&\ce{H2O2}, \ce{OH}\\
3&\ce{H2O}\\
\hline
\end{tabular}
\end{center}
\begin{center}
\captionof{table}{Volpert indices of the reaction steps in Burke2012}
\label{tab:burkevolpertrsteps}
\begin{tabular}{|l|l|}
\hline
index&reaction steps\\
\hline
0&\ce{H2 -> 2 H},
 \ce{H2 + O2 -> H + HO2},
 \ce{O2 -> 2 O}\\
\hline
1&
 \ce{H + O2 -> O + OH},
 \ce{H2 + O -> H + OH},
 \ce{HO2 + O -> O2 + OH},\\
&\ce{H + O -> OH},
 \ce{H + HO2 -> 2 OH},
 \ce{2 O -> O2},
 \ce{H + O2 -> HO2},\\
&\ce{H + HO2 -> H2 + O2},
 \ce{2 HO2 -> H2O2 + O2},
 \ce{2 H -> H2},\\
&\ce{HO2 -> H + O2},
 \ce{H2 + HO2 -> H + H2O2}\\
\hline
2&
 \ce{HO2 + OH -> H2O + O2},
 \ce{O + OH -> H + O2},
 \ce{2 OH -> H + HO2},\\
&\ce{H + OH -> H2 + O},
 \ce{OH -> H + O},
 \ce{H2O2 + OH -> H2O + HO2},\\
&\ce{H2 + OH -> H + H2O},
 \ce{H + OH -> H2O},
 \ce{2 OH -> H2O + O},\\
&\ce{H + H2O2 -> H2 + HO2},
 \ce{2 OH -> H2O2},
 \ce{H2O2 -> 2 OH},\\
&\ce{HO2 + OH -> H2O2 + O},
 \ce{H2O2 + O -> HO2 + OH},\\
&\ce{O2 + OH -> HO2 + O},
 \ce{H2O2 + O2 -> 2 HO2},\\
& \ce{H + H2O2 -> H2O + OH}\\
\hline
3&
 \ce{H2O + O2 -> HO2 + OH},
 \ce{H2O + HO2 -> H2O2 + OH},\\
&\ce{H2O + O -> 2 OH},
 \ce{H2O + OH -> H + H2O2},\\
& \ce{H + H2O -> H2 + OH},
 \ce{H2O -> H + OH}\\
\hline
\end{tabular}
\end{center}

Let us note that water only appears at the third level.
This is the same with some mechanisms in other classes:
CRECK2012 and Kereomnes2013; and in all the other mechanisms it appears
(together with all the other species and reaction steps)
earlier, at level 2.

\subsubsection{CRECK2012}
\torol{
\begin{center}
\captionof{table}{Volpert indices of the reaction steps in CRECK2012}
\label{tab:creckvolpertrsteps}
\begin{tabular}{ll}
index&reaction steps\\
\hline\\
0&\ce{O2 -> 2 O},
 \ce{H2 + O2 -> H + HO2},
 \ce{H2 -> 2 H}\\
1&
 \ce{H + O2 -> O + OH},
 \ce{HO2 + O -> O2 + OH},
 \ce{H2 + O -> H + OH},\\
 &
 \ce{H + O -> OH},
 \ce{2 O -> O2},
 \ce{H + HO2 -> 2 OH},\\
& \ce{H + HO2 -> H2 + O2},
 \ce{2 HO2 -> H2O2 + O2},
 \ce{H + O2 -> HO2},\\
& \ce{HO2 -> H + O2},
 \ce{H2 + HO2 -> H + H2O2},
 \ce{2 H -> H2}
\\
2&
 \ce{HO2 + OH -> H2O + O2},
 \ce{O + OH -> H + O2},
 \ce{O2 + OH -> HO2 + O},\\
& \ce{HO2 + OH -> H2O2 + O},
 \ce{H2O2 + O -> HO2 + OH},
 \ce{2 OH -> H2O + O},\\
& \ce{H + OH -> H2 + O},
 \ce{OH -> H + O},
 \ce{H2O2 + OH -> H2O + HO2},\\
& \ce{2 OH -> H + HO2},
 \ce{H2O2 + O2 -> 2 HO2},
 \ce{H + H2O2 -> H2O + OH},\\
& \ce{H + H2O2 -> H2 + HO2},
 \ce{2 OH -> H2O2},
 \ce{H2O2 -> 2 OH},\\
& \ce{H2 + OH -> H + H2O},
 \ce{H + OH -> H2O}
\\
3&
 \ce{H2O + O2 -> HO2 + OH},
 \ce{H2O + O -> 2 OH},
 \ce{H2O + HO2 -> H2O2 + OH},\\
& \ce{H2O + OH -> H + H2O2},
 \ce{H + H2O -> H2 + OH},
 \ce{H2O -> H + OH}
\end{tabular}
\end{center}
}

CRECK2012 contains a single \emp{irreversible step}: \ce{H2O2 + O -> HO2 + OH}.
Upon going through all the hydrogen combustion mechanisms it turns out that
no other mechanism contains any irreversible steps.
\begin{center}
\captionof{table}{Volpert indices of the species in CRECK2012}
\label{tab:creckvolpert}
\begin{tabular}{|l|l|}
\hline
index&species\\
\hline
0&\ce{H2}, \ce{O2}\\
1&\ce{H}, \ce{HO2}, \ce{O}\\
2&\ce{H2O2}, \ce{OH}\\
3&\ce{H2O}\\
\hline
\end{tabular}
\end{center}

\subsubsection{Dagaut2003}
\begin{figure}[!ht]
\begin{center}
\includegraphics[width=0.38\paperwidth]{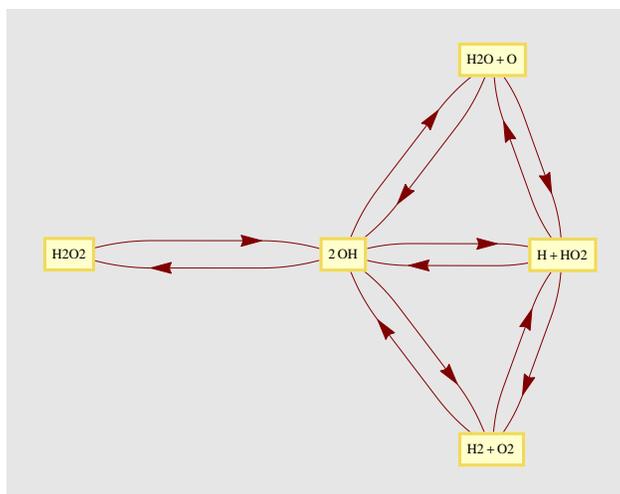}
\caption{The maximal connected component of the Feinberg--Horn--Jackson graph of Dagaut2003}
\label{fig:DagautMax}
\end{center}
\end{figure}

The maximal connected component of the \FHJg\ of Dagaut2003 is shown in Fig. \ref{fig:DagautMax}.
\begin{center}
\captionof{table}{Volpert indices of the species in Dagaut2003}
\label{tab:dagautvolpert}
\begin{tabular}{|l|l|}
\hline
index&species\\
\hline
0&\ce{H2}, \ce{O2}\\
1&\ce{H}, \ce{HO2}, \ce{O}\\
2&\ce{OHEX}, \ce{H2O2}, \ce{OH}\\
3&\ce{H2O}\\
\hline
\end{tabular}
\end{center}

\subsubsection{Keromnes2013}
\begin{figure}[!ht]
\begin{center}
\includegraphics[width=0.38\paperwidth]{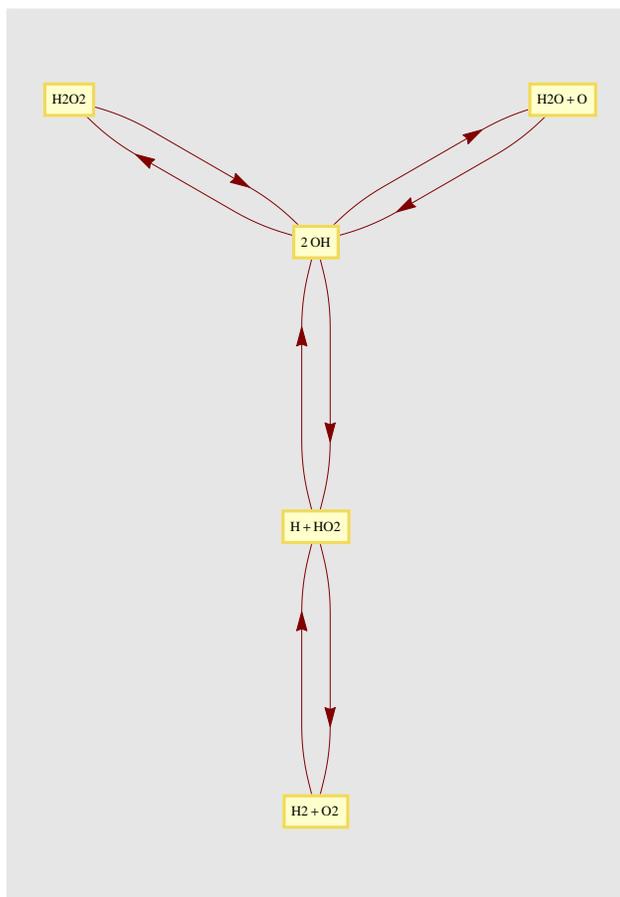}
\caption{The maximal connected component of the Feinberg--Horn--Jackson graph of Keromnes2013}
\label{fig:KeromnesMax}
\end{center}
\end{figure}

The maximal connected component of the \FHJg\ of Keromnes2013 is shown in Fig. \ref{fig:KeromnesMax}.
\begin{center}
\captionof{table}{Volpert indices of the species in Keromnes2013}
\label{tab:keromnesvolpert}
\begin{tabular}{|l|l|}
\hline
index&species\\
\hline
0&\ce{H2}, \ce{O2}\\
1&\ce{H}, \ce{HO2}, \ce{O}\\
2&\ce{H2O2}, \ce{H2O}, \ce{OH}\\
\hline
\end{tabular}
\end{center}

Another interesting application can be found. Both from the point of view of thermodynamics
and from the point of view of reducing the number of reaction rate coefficients one can require
that a mechanism be
detailed balanced, naturally, under the assumption that temperature and pressure are constant.
Applying the pair of conditions formulated by \cite{feinbergdb}
(with an appropriate numbering of reaction steps, see the Electronic Supplementary Material)
we get the following necessary and sufficient conditions in the case of Keromnes2013-type mechanisms:
\begin{eqnarray*}
&&k_{26} k_{27} k_{42}=k_{25} k_{28} k_{41},\quad
k_4 k_{13} k_{39}=k_3k_{14} k_{40},\quad
k_2 k_{14} k_{17}=k_1 k_{13} k_{18},\\
&&k_3 k_8 k_{11}=k_4 k_7k_{12},\quad
k_2 k_4 k_5=k_1 k_3 k_6,\quad
k_2 k_4 k_9 k_{21}=k_1 k_3 k_{10}k_{22},\\
&&k_2 k_8 k_9 k_{19}=k_1 k_7 k_{10} k_{20},\quad
k_2 k_9 k_{14} k_{15}=k_1k_{10} k_{13} k_{16},\\
&&k_2 k_8 k_9 k_{13} k_{23} k_{37}^2=k_1 k_7 k_{10}k_{14} k_{24} k_{38}^2,\quad
k_2 k_8 k_9 k_{14} k_{24} k_{35}^2=k_1 k_7 k_{10}k_{13} k_{23} k_{36}^2,\\
&&k_1 k_8 k_{10} k_{13} k_{24} k_{33}^2=k_2 k_7 k_9k_{14} k_{23} k_{34}^2,\quad
k_1 k_8 k_9 k_{13} k_{24} k_{29}^2=k_2 k_7 k_{10}k_{14} k_{23} k_{30}^2,\\
&&k_2 k_8 k_9 k_{13} k_{24} k_{25}^2=k_1 k_7 k_{10}k_{14} k_{23} k_{26}^2,\quad
k_2 k_4^2 k_8 k_9 k_{13} k_{24} k_{31}^2=k_1 k_3^2k_7 k_{10} k_{14} k_{23} k_{32}^2
\end{eqnarray*}
Further investigations may also use the extended theory by \cite{gorbanyablonsky}.

\subsubsection{Starik2009}
\begin{center}
\captionof{table}{Volpert indices of the species in Starik20009}
\label{tab:starikvolpert}
\begin{tabular}{|l|l|}
\hline
index&species\\
\hline
0&\ce{H2}, \ce{O2}\\
1&\ce{H}, \ce{HO2}, \ce{O3}, \ce{O}\\
2&\ce{H2O2}, \ce{H20}, \ce{OH}\\
\hline
\end{tabular}
\end{center}

\begin{figure}[!hb]
\begin{center}
\includegraphics{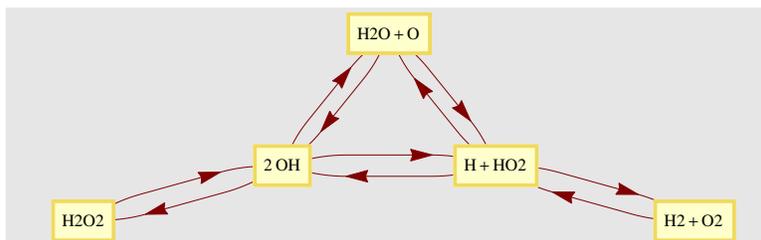}
\caption{The maximal connected component of the Feinberg--Horn--Jackson graph of Starik2009}
\label{fig:StarikMax}
\end{center}
\end{figure}

The maximal connected component of the \FHJg\ of Starik2009 is shown in Fig. \ref{fig:StarikMax}.

\subsubsection{Similarities and differences between the mechanisms of hydrogen combustion}
We can further analyse the similarities and differences between the class representatives
of hydrogen combustion mechanisms. (Similarity of mechanisms is understood in the na\"{\i}ve way:
the more common species and reaction steps two underlying reactions contain the more similar they are.)
One can easily determine which are the reaction steps present in one mechanism and
missing in the other.
This generates a very huge table, a simpler one is obtained
if one only counts the number of reaction steps (reversible counts 2)
present in one mechanism and missing from the other.

\begin{center}
\captionof{table}{Number of different reaction steps in different mechanisms I.}
\label{tab:hdifferent1}
\begin{tabular}{|l|c|c|c|c|}
\hline
&Ahmed&Burke&CRECK&Dagaut\\
&2007&2012&2012&2003\\
\hline
Ahmed2007&  0 & 2 & 3 & 0\\
Burke2012&  2 & 0 & 3 & 0\\
CRECK2012& 2 & 2 & 0 & 2\\
Dagaut2003& 4 & 4 & 7 & 0\\
Davis2005& 2 & 2 & 5 & 0\\
Keromnes2013& 6 & 4 & 7 & 4\\
SanDiego2011&  4 & 4 & 5 & 2\\
Starik2009& 14 & 14 & 17 & 12\\
\hline
\end{tabular}
\end{center}
\begin{center}
\captionof{table}{Number of different reaction steps in different mechanisms II.}
\label{tab:hdifferent2}
\begin{tabular}{|l|c|c|c|c|}
\hline
&Davis&Keromnes&SanDiego&Starik\\
&2005&2013&2011&2009\\
\hline
Ahmed2007&   0 & 2 & 0 & 0 \\
Burke2012&   0 & 0 & 0 & 0 \\
CRECK2012&   2 & 2 & 0 & 2 \\
Dagaut2003&  2 & 4 & 2 & 2 \\
Davis2005&   0 & 2 & 0 & 0 \\
Keromnes2013&  4 & 0 & 4 & 4 \\
SanDiego2011&  2 & 4 & 0 & 2 \\
Starik2009&    12 & 14 & 12 & 0 \\
\hline
\end{tabular}
\end{center}

Tables \ref{tab:hdifferent1} and \ref{tab:hdifferent2} shows the number of reaction steps missing in the respective "column" mechanism, but present in the "row" mechanism.

Let us look at a single example. The reaction steps included in Ahmed2007, but not in Burke2012, CRECK2012 and Keromnes2013 are \ce{H + HO2 <=> H2O + O} in all cases, and
in the case of CRECK2012 \ce{H2O2 + O <- HO2 + OH}, as well.
Reaction steps contained in CRECK2012, but missing in Ahmed2007 are
\ce{HO2 <=> O + OH}.
This shows that in Tables \ref{tab:hdifferent1} and \ref{tab:hdifferent2} reaction steps may mean either an irreversible step, or a reversible pair. Furthermore, Table \ref{tab:hdifferent2} shows again that Starik2009 and Keromnes2013 contain  quite a few reaction steps missing in the other mechanisms.

Readers interested in detailed combustion chemistry are referred to the supplement.

\subsection{Carbon monoxide}
Let us now summarize the basic data of carbon monoxide combustion mechanisms in
Table \ref{tab:cobasic}.

\begin{center}
\captionof{table}{Basic data of the investigated carbon monoxide combustion mechanisms}
\label{tab:cobasic}
\begin{tabular}{|l|l|l|l|l|}
\hline
{\bf Mechanism}&{\bf Reference}&$M$&$R$&$\delta=N-L-S$\\
\hline
Ahmed2007&\cite{ahmedmaussmoreaczeuch}&
12&72&$57-23-9=25$\\
\hline
CRECK2012&\cite{healykalitanaulpetersenbourquecurran}&
11&60&$49-19-8=22$\\
\hline
Dagaut2003&\cite{dagautlecomtemieritzglarborg}&
12&68&$52-21-9=22$\\
\hline
Davis2005&\cite{davisjoshiwangegolfopoulos}&
11&60&$47-19-8=20$\\
\hline
GRI30&\cite{smithgoldenfrenklachmoriaryeiteneergoldenbergbowmanhansonsonggardinerlissianski}&
12&74&$57-23-9=25$\\
\hline
Keromnes2013&\cite{keromnesmetcalfeheuferdonohoedassungherzlernaumanngriebelmathieukrejcipetersenpitzcurran}&
12&64&$52-21-9=22$\\
\hline
Li2007&\cite{lizhaokazakovchaosdryerscire}&
12&78&$61-24-9=28$\\
\hline
NUIG2010&\cite{healykalitanaulpetersenbourquecurran}&
12&78&$61-24-9=28$\\
\hline
Rasmussen2008&\cite{rasmussenhansenmarshallglarborg}&
13&88&$66-26-10=30$\\
\hline
SanDiego2011&\cite{sandiego}&
12&74&$57-23-9=25$\\
\hline
SaxenaWilliams2006&\cite{saxenawilliams}&
11&60&$45-18-8=19$\\
\hline
Starik2009&\cite{stariktitovasharipovkozlov}&
13&88&$70-28-10=32$\\
\hline
Sun2007&\cite{sunyangjomaaslaw}&
12&66&$52-21-9=22$\\
\hline
USC2007&\cite{wangyoujoshidavislaskinegolfopouloslaw}&
12&74&$57-23-9=25$\\
\hline
Zsely2005&\cite{zselyzadorturanyi}&
11&62&$47-19-8=20$\\
\hline
\end{tabular}
\end{center}
Since all CO and hydrocarbon mechanisms contain a subset of hydrogen reactions, they also appear in Table \ref{tab:hbasic1} \& \ref{tab:hbasic2}. Hydrogen-only mechanisms (OConaire2004, Konnov2008, Hong2011, Burke2012) were not used here.
Although the references show a large overlap with those in Table \ref{tab:hbasic1} \& \ref{tab:hbasic2},
here we focus on the submechanism describing carbon monoxide combustion.
Now let us start finding the reasons why we have different numbers in different mechanisms.
\begin{description}
\item[Species and their number, $M$]
Here, the values are much more diverse. The core species present in all the mechanisms are
\ce{CO}, \ce{CO2}, \ce{H}, \ce{H2}, \ce{H2O}, \ce{HCO}, \ce{HO2},
\ce{H2O2}, \ce{O}, \ce{O2}, \ce{OH}.
Additional species are listed in  \ref{tab:cospecies}.
\begin{center}
\captionof{table}{Species present only in a given mechanism}
\label{tab:cospecies}
\begin{tabular}{|l|l|}
\hline
Particular species&Mechanisms\\
\hline
\ce{CH2O}&Ahmed2007, Dagaut2003, GRI30, Li2007, \\
&NUIG2010, Rasmussen2008, SanDiego2011\\
&Starik2009, Sun2007, USC2007\\
\hline
\ce{OHEX}&Keromnes2013\\
\hline
\ce{HOCO}&Rasmussen2008\\
\hline
\ce{O3}&Starik2009\\
\hline
\end{tabular}
\end{center}
\item[Number of reaction steps, $R$]
The number of reaction steps, $R$, varies between 60 and 88.
The law of C. K. Law reported
in his comprehensive review paper
\citep{law} that the number of reactions is approximately 5 times larger than
the number of species is fulfilled again.
\item[Deficiency, $\delta$]
The number of complexes, $N$, varies between 45 and 66.
The number of weakly connected components is between 18 and 28.
The deficiencies are large, neither the zero deficiency theorem,
nor the one deficiency theory can be applied.
\item[Weak reversibility and acyclicity]
None of the reactions have an acyclic Volpert graph, as all the reactions,
except CRECK2012 and Keromnes2013,
are fully reversible.
Both the mentioned two mechanisms contain two irreversible steps:
$$
\ce{HCO + HO2 -> CO2 + H + OH} \mathrm{\ , \ } \ce{H2O2 + O -> HO2 + OH} \mathrm{\ (CRECK2012);}
$$
$$
\ce{2 HCO -> 2 CO + H2} \mathrm{\ , \ } \ce{HCO + HO2 -> CO2 + H + OH} \mathrm{\  (Keromnes2013).}
$$
Although irreversible steps are acceptable modelling tools,
there are at least two problems with them.
If only one direction of the reaction is used the negligibility
of the reverse reaction step may depend on the circumstances and
it is possible that the mechanism will be used at such conditions
where this simplification assumption will not be valid.
In case both directions are present in a mechanism both of their
values should change according to the thermodynamic equilibrium
if they are re-parametrized.
\end{description}

\subsubsection{Representations of mechanism classes}
The situation is much simpler here, in the case of carbon monoxide combustion mechanisms.
From the structural point of view USC2007 is identical to GRI30, while
NUIG2010 is identical to Li2007, and all the other mechanisms are different.
Hence we do not introduce classes of mechanisms in this case.

\subsubsection{Similarities and differences between the mechanisms of carbon monoxide combustion}
\begin{center}
\captionof{table}{Species present in one CO mechanism and missing in others 1}
\label{tab:cospe1}
\begin{tabular}{|l|l|l|l|l|l|l|l|}
\hline
Mechanism&Ahmed&CRECK&Dagaut&Davis&GRI30&Keromnes&Li  \\
         &2007 &2012 &2003  &2005 &     &2013    &2007\\
\hline
Ahmed2007& 0 & 19 & 12 & 14 & 2 & 16 & 2 \\
\hline
CRECK2012&7 & 0 & 6 & 7 & 7 & 6 & 6 \\
\hline
Dagaut2003&8 & 14 & 0 & 8 & 8 & 8 & 4 \\
\hline
Davis2005&2 & 7 & 0 & 0 & 0 & 2 & 2 \\
\hline
GRI30& 4 & 21 & 14 & 14 & 0 & 16 & 2 \\
\hline
Keromnes&8 & 10 & 4 & 6 & 6 & 0 & 4 \\
2013&&&&&&&\\
\hline
Li2007&8 & 24 & 14 & 20 & 6 & 18 & 0 \\
\hline
Rasmussen&16 & 34 & 24 & 28 & 14 & 29 & 12 \\
2008&&&&&&&\\
\hline
SanDiego& 4 & 19 & 14 & 14 & 2 & 16 & 4 \\
2011&&&&&&&\\
\hline
Saxena& 4 & 7 & 2 & 2 & 2 & 4 & 4 \\
Williams&&&&&&&\\
2006&&&&&&&\\
\hline
Starik2009& 16 & 35 & 24 & 28 & 16 & 29 & 14 \\
\hline
Sun2007& 6 & 12 & 0 & 6 & 6 & 6 & 2 \\
\hline
Zsely2005& 4 & 9 & 0 & 2 & 2 & 4 & 4 \\
\hline
\end{tabular}
\end{center}
Tables \ref{tab:cospe1} and \ref{tab:cospe2} shows the number of species missing in the respective "column" mechanism, but present in the "row" mechanism.
\begin{center}
\captionof{table}{Species present in one CO mechanism and missing in others 2}
\label{tab:cospe2}
\begin{tabular}{|l|l|l|l|l|l|l|}
\hline
Mechanism&Rasmussen&SanDiego&Saxena&Starik&Sun&Zsely  \\
         &         &        &Williams&    &    &\\
         &2008     &2011    &2006   &2009 &2007&2005\\
\hline
Ahmed2007&  0 & 2 & 16 & 0 & 12 & 14 \\
\hline
CRECK2012&6 & 5 & 7 & 7 & 6 & 7 \\
\hline
Dagaut2003& 4 & 8 & 10 & 4 & 2 & 6 \\
\hline
Davis2005& 0 & 0 & 2 & 0 & 0 & 0 \\
\hline
GRI30& 0 & 2 & 16 & 2 & 14 & 14 \\
\hline
Keromnes&5 & 6 & 8 & 5 & 4 & 6 \\
2013&&&&&&\\
\hline
Li2007& 2 & 8 & 22 & 4 & 14 & 20 \\
\hline
Rasmussen& 0 & 16 & 30 & 14 & 24 & 28 \\
2008&&&&&&\\
\hline
SanDiego2011& 2 & 0 & 14 & 2 & 14 & 14 \\
\hline
Saxena& 2 & 0 & 0 & 2 & 2 & 2 \\
Williams&&&&&&\\
2006&&&&&&\\
\hline
Starik2009&  14 & 16 & 30 & 0 & 24 & 28 \\
\hline
Sun2007& 2 & 6 & 8 & 2 & 0 & 6 \\
\hline
Zsely2005& 2 & 2 & 4 & 2 & 2 & 0 \\
\hline
\end{tabular}
\end{center}
As an illustration let us calculate the Volpert indices of Zsely2005
under the assumption that the species \ce{O2}, \ce{H2} and \ce{CO}
are initially present.
\begin{center}
\captionof{table}{Volpert indices of the species in Zsely2005}
\label{tab:zselyvolpert}
\begin{tabular}{|l|l|}
\hline
index&species\\
\hline
0&\ce{H2}, \ce{O2}, \ce{CO}\\
1&\ce{H}, \ce{O}, \ce{HO2}, \ce{OH}, \ce{HCO}, \ce{CO2}\\
2&\ce{H2O2},  \ce{H2O}\\
\hline
\end{tabular}
\end{center}

The maximal (weakly) connected components of the Feinberg--Horn--Jackson graphs
are the same as those found in the case of hydrogen combustion mechanisms.
The reason for this is that there are not enough carbon
containing species in the mechanisms to form larger components,
which is not the case with methanol mechanisms.

\subsection{Methanol}
Let us start again with the basic data, see Table \ref{tab:mbasic}.

\begin{center}
\captionof{table}{Basic data of the investigated methanol combustion mechanisms}
\label{tab:mbasic}
\begin{tabular}{|l|l|l|l|l|}
\hline
{\bf Mechanism}&{\bf Reference}&$M$&$R$&$\delta=N-L-S$\\
\hline
Aranda2013&\cite{arandachristensenalzuetaglarborggersencgaodmarshall}&
76&1063&$661-187-71=403$\\
\hline
Klippenstein2011&\cite{klippensteinhardingdavistomlinskodje}&
18&172&$122-42-15=65$\\
\hline
Li2007&\cite{lizhaokazakovchaosdryerscire}&
18&170&$121-42-15=64$\\
\hline
Rasmussen2008&\cite{rasmussenwassarddamjohansenglarborg}&
28&320&$222-75-24=123$\\
\hline
ZabettaHupa2008&\cite{zabettahupa}&
58&724&$500-163-54=283$\\
\hline
\end{tabular}
\end{center}
The analysis of these mechanisms is much harder.
\begin{description}
\item[Species, classes of mechanisms]
The number of reaction steps is around ten times that of the species here.

There is a striking similarity of Klippenstein2011 and Li2007 at the level of numbers.
Really, they use the same set of species, and the only difference between
their reaction steps is that Klippenstein2011
contains also the reversible reaction step
$$
\ce{CH3O + H2O2 <=> CH3OH + HO2}
$$
in addition to the common reaction steps.
It is in accordance with the statement of the authors that they
only made a small change on the structure of Li2007.

Otherwise, methanol mechanisms are so different that
the question of classes and their representation does not even come up.
\item[Number of reaction steps, $R$]
The number of reaction steps, $R$, ranges between 18 and 76.
\item[Weak reversibility and acyclicity]
None of the reactions have an acyclic Volpert graph, as Li2007, Rasmussen2008 and ZabettaHupa2008
are fully reversible, and most of the reaction steps of the two other reactions
are reversible. The exceptions are that
Aranda2013 contains the irreversible reaction steps
\begin{eqnarray*}
&\ce{CH2OOH ->  CH2O + OH}, &\\
&\ce{C2H4 + HOCH2CH2OO ->  CH2O + CH2OH + CH3CHO},&\\
&\ce{CH2O + HOCH2CH2OO ->  CH2OH + CH2OOH + HCO},&\\
&\ce{HO2 + HOCH2CH2OO ->  CH2OH + CH2OOH + O2},&\\
&\ce{CH2CHOH + O2 ->  CH2O + HCO + OH};&
\end{eqnarray*}
whereas Rasmussen2008
contains the irreversible reaction steps
$$
\ce{NO2 ->  NO2*}, \ce{2 NO2* ->  2 NO + O2},
$$
and all the other reaction steps in all the other mechanisms are reversible.
\item[Deficiency, $\delta$]
As even the smallest \FHJg\ is too large to be shown here,
we shall again deal with the largest components of the \FHJg\ of the
individual mechanisms. Volpert graphs will only be used for indexing,
and will show some interesting relationships.
\end{description}

\subsubsection{Aranda2013}
\begin{figure}[!htb]
\begin{center}
\includegraphics[width=0.54\paperwidth]{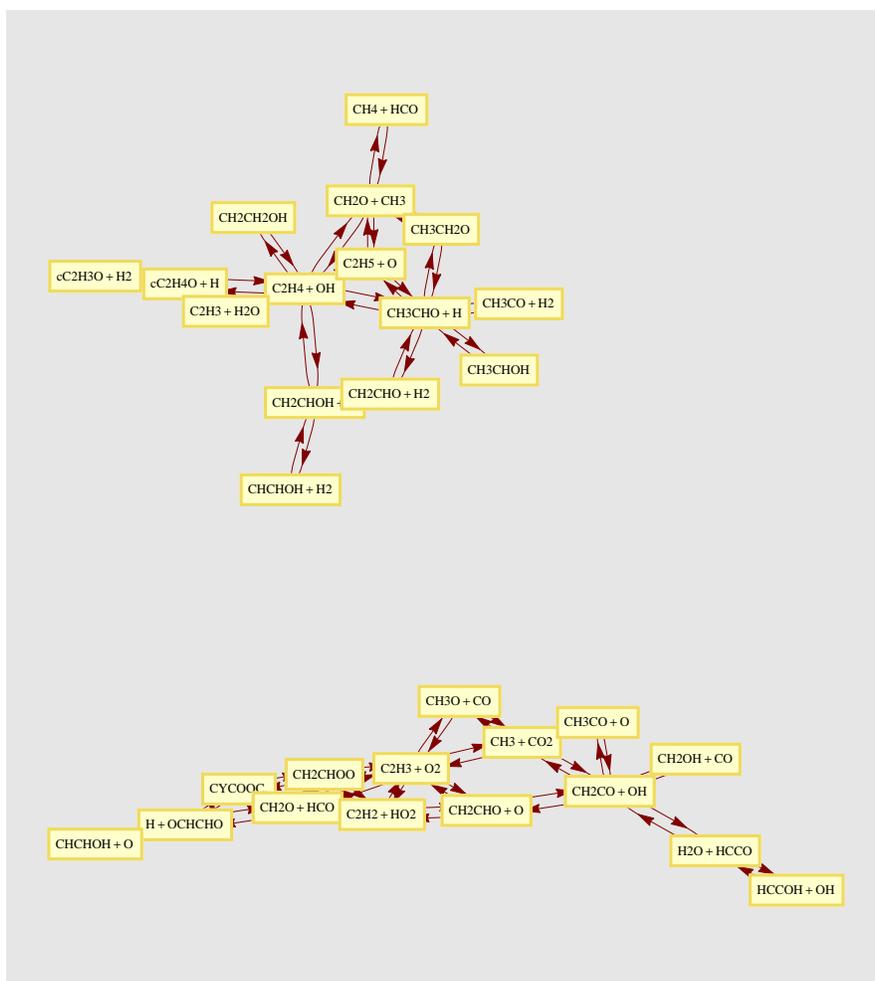}
\caption{The maximal weakly connected components of Aranda2013}
\label{fig:ArandaMaxComp}
\end{center}
\end{figure}

The maximal weakly connected components of the \FHJg\ of Aranda2013 are shown in Fig. \ref{fig:ArandaMaxComp}.

Starting from \ce{CH2O} as initial species all the reaction steps can finally take place and
all the species will be produced, except those reaction steps where compounds of nitrogen occur
in the reactant complex. If one takes
$\{\ce{CH2O}, \ce{NO2}, \ce{NH3}\}$
as the initial set,
then all the reaction steps are capable of taking place and all the species will be produced
and the largest Volpert index is now 4.\\
There are 22 species containing two carbon atoms (or C--C bonds, as these expressions are
synonymous in this case):
\ce{HOCH2CH2OO}, \ce{CH3CHO}, \ce{CH2CHOH}, \ce{H2CC}, \ce{CH2CHO},
\ce{CH2CO}, \ce{CH3CO}, \ce{CH2CH2OH}, \ce{CH2CH2OOH}, \ce{CH3CH2OO}, \ce{CHCHOH},
\ce{CH2CHOO}, \ce{CYCOOC.}, \ce{CH2CHOOH}, \ce{CH3CH2O}, \ce{CH3CH2OH},
\ce{CH3CHOH}, \ce{HCCO}, \ce{CH3CH2OOH}, \ce{CH3CHOOH}, \ce{HCCOH} and \ce{OCHCHO}.\\
Beyond Aranda2013 it is only ZabettaHupa2008 where species with not less than
two carbon atoms can be found.

\subsubsection{Klippenstein2011 and Li2007}
\begin{figure}[!htb]
\begin{center}
\includegraphics[width=0.5\paperwidth]{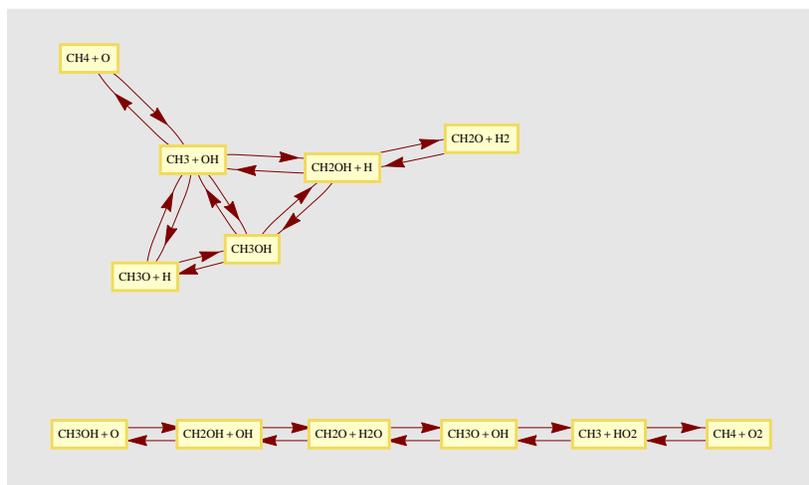}
\caption{The maximal weakly connected components of Klippenstein2011 and Li2007}
\label{fig:KlippensteinAndLiMaxComp}
\end{center}
\end{figure}

The maximal weakly connected components of the \FHJg\ of Klippenstein2011
or Li2007 are shown in Fig. \ref{fig:KlippensteinAndLiMaxComp}.

Starting from \ce{CH2O} as initial species all the reaction steps can finally take place and
all the species will be produced. The species \ce{H2O2} and all the reaction steps where
this species is a reactant species (and only those) will appear latest, only at level four.

\subsubsection{Rasmussen2008}
\begin{figure}[!htb]
\begin{center}
\includegraphics[width=0.43\paperwidth]{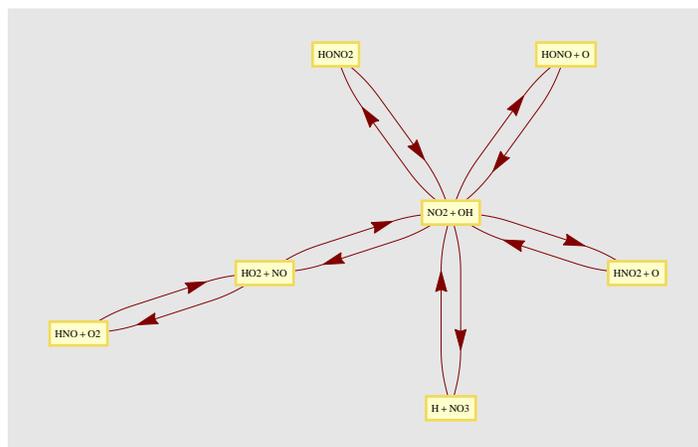}
\caption{The maximal weakly connected component of Rasmussen2008}
\label{fig:RasmussenMaxComp}
\end{center}
\end{figure}

The maximal weakly connected component of the \FHJg\ of Rassmussen2008 is shown
in Fig. \ref{fig:RasmussenMaxComp}.

Starting from \ce{CH2O} as initial species all the reaction steps can finally take place and
all the species will be produced, except those reaction steps where compounds of nitrogen occur
in the reactant complex. If one also adds \ce{NO} initially, then all the reaction steps
are capable of taking place and all the species will be produced.
If the set of initial species is $\{\ce{CH2O}, \ce{NO2}\}$, then the situation is even better: 
the largest index is now only 3. 
Let us emphasize that the statements of this paragraph (and similar statements below)
are independent from the values of the reaction rate coefficients.

\subsubsection{ZabettaHupa2008}
\begin{figure}[!htb]
\begin{center}
\includegraphics[width=0.5\paperwidth]{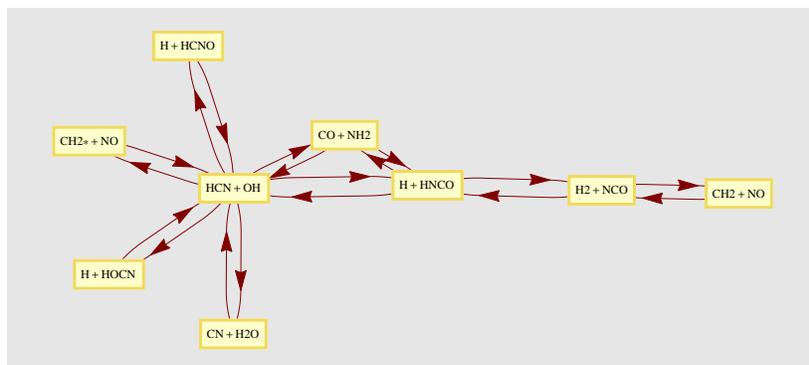}
\caption{The maximal connected component of ZabettaHupa2008}
\label{fig:ZabettaHubaMaxComp}
\end{center}
\end{figure}

The maximal connected component of the \FHJg\ of ZabettaHupa2008 is shown
in Fig. \ref{fig:ZabettaHubaMaxComp}.

Again, if the initial species is only \ce{CH2O}, then no nitrogen compounds (and radicals etc.)
are produced. However, adding \ce{NH3} to the initial list of species, all the species and reaction steps
receive a finite Volpert index. 
Species containing two carbon atoms are \ce{CH2CO}, \ce{HCCO}.

\subsubsection{Similarities and differences between the mechanisms of methanol combustion}
The mechanisms are too large (especially Aranda2013) to present all details here.
However, it is possible to show the number of reaction steps present in the different
mechanisms and missing in the others, see Table \ref{tab:mdifferences}.
\begin{center}
\captionof{table}{The number of different reaction steps in the different mechanisms}
\label{tab:mdifferences}
\begin{tabular}{|l|r|r|r|r|r|}
\hline
&Aranda&Klippenstein&Li&Rasmussen&ZabettaHupa\\
&2013&2011&2007&2008&2008\\
\hline
Aranda2013& 0 & 897 & 899 & 759 & 737 \\
Klippenstein2011& 6 & 0 & 2 & 8 & 18 \\
Li2007& 6 & 0 & 0 & 6 & 16 \\
Rasmussen2008& 16 & 156 & 156 & 0 & 120 \\
ZabettaHupa2008& 398 & 570 & 570 & 524 & 0 \\
\hline
\end{tabular}
\end{center}
Thus, 897 in the first row, second column in Table \ref{tab:mdifferences} means
that there are altogether 897 reaction steps  enumerated in Aranda2013 but missing
in Klippenstein2011. Note that the table is not symmetric, it should not be in general.

Let us look at a few examples in more detail.
As Li2007 is a proper subset of Klippenstein2011,
there is no reaction step present in the first one and missing in the second.
The reaction steps present in Klippenstein2011 and missing in ZabettaHupa2008 are:
\begin{eqnarray*}
&\ce{2 CH2O <=> CH2OH + HCO},\quad
 \ce{CH2O + CO <=> 2 HCO}, &\\
& \ce{CH3O + CO <=> CH3 + CO2},\quad
\ce{CH3O + H <=> CH3 + OH},  &\\
& \ce{2 CO + H2 <=> 2 HCO},\quad
 \ce{CH2O + H2O2 <=> CH2OH + HO2},&\\
&\ce{CH2O + H2O2 <=> CH3O + HO2},\quad
 \ce{CH3O + H2O2 <=> CH3OH + HO2}, &\\
& \ce{HCO + HO2 <=> CO2 + H + OH}&
\end{eqnarray*}
Finally, Rasmussen2008 contains a few reaction steps among nitrogen compounds
(including two irreversible steps)
which are not present in the huge Aranda2013. These are as follows.
\begin{eqnarray*}
&\ce{CH2O + H2O2 <=> CH3O + HO2},\quad
 \ce{HNO2 <=> H + NO2},& \\
&\ce{ HO2 + NO2 <=> NO3 + OH},\quad
\ce{NO3 <=> NO + O2},  &\\
&\ce{ H + NO3 <=> NO2 + OH},\quad
 \ce{HO2 + NO3 <=> NO2 + O2 + OH},& \\
&\ce{NO3 + O <=> NO2 + O2},\quad\ce{NO2 -> NO2*},\quad
 \ce{2 NO2* -> 2 NO + O2}&\\
 \end{eqnarray*}

\section{Discussion and outlook}
The major application of the methods outlined in the paper
is a structural analysis of the selected mechanisms prior to a
quantitative analysis including the evaluation of reaction rate constants.

A systematic use of Volpert indexing may also serve the selection of
a minimal initial set of species: the least number of species which is enough
for all the reaction steps in a given mechanism to occur and for all the species to be produced.

Another possible application is that one starts from a big mechanism
and deletes reaction steps obeying some restrictions. E.g. one starts
from a CO combustion mechanism and deletes reaction steps containing C,
thus we should arrive at a hydrogen combustion mechanism etc.
The results are only useful if they are exported to a CHEMKIN file,
\mma{CHEMKINExport} will serve for this purpose.

Additional fields of application of our method are metabolism chemistry as well as atmospheric chemistry.
\section*{Acknowledgments}
The authors thank the cooperation with Prof. Tam\'as Tur\'anyi,
Mr. Carsten Olm and Mr. R\'obert P\'alv\"olgyi.
Ms. \'Agota Busai was so kind as to closely read earlier versions of the manuscript.
Many of the participants of MaCKiE 2013 contributed with incentive ideas.
Further requirements, criticism and problems to be solved are wanted.
\section*{Appendix: Fundamentals for Formal Kinetics}\label{subsec:graphs}
The basic notions can be found in textbooks such as \cite{erditoth,feinbergces1,feinbergces2,volperthudyaev,yablonskimarin} etc.
\torol{
\subsection{Feinberg--Horn--Jackson graphs, Volpert graphs, Volpert indices of reactions}
}
Let us consider the reaction
\begin{eqnarray}\label{reaction}
\sum_{m=1}^{M}\alpha(m,r)X(m)\longrightarrow\sum_{m=1}^{M}\beta(m,r)X(m)\quad
(r=1,2,\dots,R)
\end{eqnarray}
with $M\in\N$ chemical \emp{species}: $X(1),X(2),\dots,X(M);{\ }
R\in\N$  \emp{reaction steps},
$$
\alpha(m,r),\ \beta(m,r)\in\N_0{\ }
(m=1,2,\dots,M; r=1,2,\dots,R)
$$
\emp{stoichiometric coefficients} or \emp{molecularities}.
Mind that we count a reversible reaction step as two reaction steps, although chemists count it as one, sometimes.\\
Furthermore, suppose the deterministic model of \eqref{reaction} is
\begin{eqnarray}\label{ikde}
\dot{c}_m(t)&=&f_m(\cb(t)):=
\sum_{r=1}^R(\beta(m,r)-\alpha(m,r))w_{r}(\cb(t))\\
c_m(0)&=&c_{m}^{0}\in\R_0^+\quad(m=1,2,\dots,M),\label{ini}
\end{eqnarray}
describing the time evolution of the concentration vs. time functions
$
t\mapsto c_m(t):=[X(m)](t)
$
of the species, which is based on \emp{mass action-type kinetics}:
\begin{equation*}
w_{r}(\overline{\cb}):=k_{r}\overline{\cb}^{\alpha(.,r)}:=
k_{r}\prod_{p=1}^{M}\overline{\cb}_{p}^{\alpha(p,r)}
(r=1,2,\dots,R),
\end{equation*}
where the constants $k_r\in\R$ are referred to as the \emp{reaction rate coefficients}. \eqref{ikde} is also called the \emp{(induced) kinetic differential equation} of the reaction \eqref{reaction} (see \cite{erditoth}).

When reaction \eqref{reaction} is \emp{reversible} one can ask that whether the molecular process is \emp{detailed balanced} or not at equilibrium. It means that a reaction step and its reverse occur, on the average, at the same rate. More formally, equations $k_r(\cb^*)^{\alpha(\cdot,r)} = k_{-r}(\cb^*)^{\beta(\cdot,r)}$ are required to be satisfied for equilibrium points $\cb^*$ and for all the reversible reaction step pairs, where $k_r$, $k_{-r}$ denote the  corresponding reaction rate coefficients (see \cite{nagypapptoth,nagytoth} and references therein)

The number of \emp{complexes} $N$ is the number of different
\emp{complex vectors} among $\alpha(.,r)$ and $\beta(.,r)$ ($r\in\{1,2,\ldots,R\}$).
The \emp{Feinberg--Horn--Jackson graph} (or, FHJ graph, for short) of the reaction is a directed graph obtained if one writes down all the complex vectors (or simply the \emp{complexes})
exactly once and connects two complexes with a directed edge (or two
different edges pointing into opposite directions)
if the first one is transformed into the second by a
reaction step. We note that the FHJ graph is a useful tool to decide
with the method proposed by \cite{feinbergdb} whether a reversible reaction is \emp{detailed balanced} or not.

A subgraph $H$ of the FHJ graph is called \emp{(strongly) connected} if between any of its two vertices (complexes) there is a directed (reaction) path: a sequence of concatenated reaction steps. We say that a subgraph $H$ of the FHJ graph is \emp{weakly connected} if between any two vertices (complexes) of the undirected version of $H$ there is a path. The maximal weakly (strongly) connected subgraphs of the FHJ graph are called the \emp{weakly (strongly) connected components}. Notice that a strongly connected subgraph is automatically weakly connected as well, but the converse is not true in general. For reversible reactions the concepts coincide. Furthermore, the weakly connected components are also called as \emp{linkage classes} in the literature. The number of weakly connected components of the FHJ graph is denoted by $L$. Finally, the \emp{maximal weakly (strongly) connected component} of the FHJ graph is any of the weakly (strongly) connected components
with the maximal number of vertices.

The \emp{stoichiometric space} is the linear subspace of $\R^M$ generated by the \emp{reaction vectors}:
$\{\beta(.,r)-\alpha(.,r); r\in\{1,2,\dots,R\}\}$; its dimension is denoted by $S$. 
Finally, the nonnegative integer
$\delta:=N-(L+S)$ is the \emp{deficiency} of the reaction \eqref{reaction}. 

\torol{In plain words, the deficiency is a non-negative integer which measures how many more different complexes are presented as many as different ways the reaction steps can transform them from one to an other involved in reaction \eqref{reaction}, which can be measured in two sense: algebraically ($S$) and structurally ($L$).}

The larger the deficiency is the more richer the mechanism is in complexes. 
We note that the notion of the deficiency plays an important role in characterizations and stability of the deterministic model of reaction \eqref{reaction} (see \cite{erditoth} and references therein, the papers by Feinberg in this journal or the recent papers  \cite{borossingle,borosmulti,borosexi}.

The \emp{Volpert graph} of the reaction is a directed bipartite graph, its two vertex sets are the species set and the set of reaction steps, and an arrow is drawn from species
$X(m)$ to the reaction step $r$ if $\alpha(m,r)>0;$ (species $X(m)$ is needed
to the reaction step $r$ to take place)
and an arrow goes from reaction $r$ to species $X(m)$  if $\beta(m,r)>0$ (species $X(m)$ is produced in the reaction step $r$). Sometimes it is worth labeling the edges with $\alpha(m,r)$ and $\beta(m,r),$ respectively.

It is very useful to assign indices to the vertices of the Volpert graph.
This goes in the following way. A subset of species is selected,
this will be the \emp{initial set}. In real applications this
will be the set of species with positive initial concentrations.
The elements of the initial set receive index zero together
with all the reaction steps which can proceed once the
\emp{initial species} are present.
Next, species without an index which can be produced by the
indexed reaction steps receive the index one, and reaction steps without
index which can proceed receive also one, and so on.
As the Volpert graph is finite, the procedure finishes in a finite number of steps.
At the end either all the vertices receive an index, either a finite value $\kappa$
or the infinite index $\infty$.
One of the many possible interpretations of the meaning of a finite index
$\kappa$ is that the given species or reaction step can only appear in the
$\kappa$th step or at the $\kappa$th level.
In accordance with this, species with an infinite index cannot be produced,
reaction steps with an infinite index cannot proceed with the prescribed
initial species of the reaction.
These statements and some others not less important ones can be found in a precise form
e.g. in \cite{volperthudyaev} or in the original paper \cite{volpert}.
An application of the Volpert index in the decomposition of overall reactions is given in \cite{kovacsvizvaririedeltoth}.

Simple examples from combustion theory follow to illustrate the meaning of the definitions.

\begin{Exa}[Mole reaction]
The earliest combustion mechanism (which has been given a detailed treatment from the point of view of the qualitative theory of differential equations) is probably the Mole reaction (\cite{mole}), see also Fig. \ref{fig:MoleVgraph}:
\begin{equation}\label{eq:fhjmole}
  \ce{Y <=> 0 <=> X}\quad\ce{X + Y -> 2X + 2Y}.
\end{equation}
\begin{figure}[!hb]
\begin{center}
\includegraphics{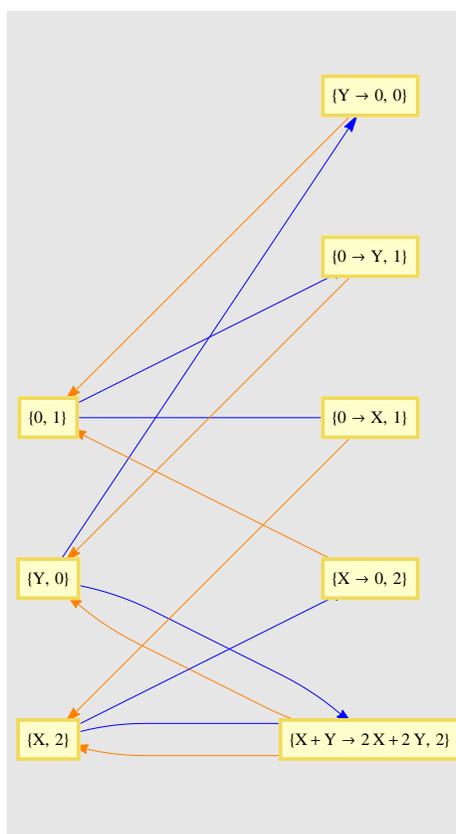}
\caption{The Volpert graph of the Mole reaction \eqref{eq:fhjmole}}
\label{fig:MoleVgraph}
\end{center}
\end{figure}
\end{Exa}

\begin{Exa}[Robertson reaction]
The reaction proposed in \cite{robertson} contains three species, its \FHJg\ is
\begin{equation}\label{eq:fhjrob}
  \ce{A -> B}\quad \ce{2B -> B + C -> A + C},
\end{equation}
the complexes are A, B, 2B, B + C, A + C, the deficiency is $N-L-S=5-2-2=1.$
The Volpert graph of this reaction is shown in Fig. \ref{fig:RobVgraph}.
\begin{figure}[!hb]
\begin{center}
\includegraphics[width=0.26\paperwidth]{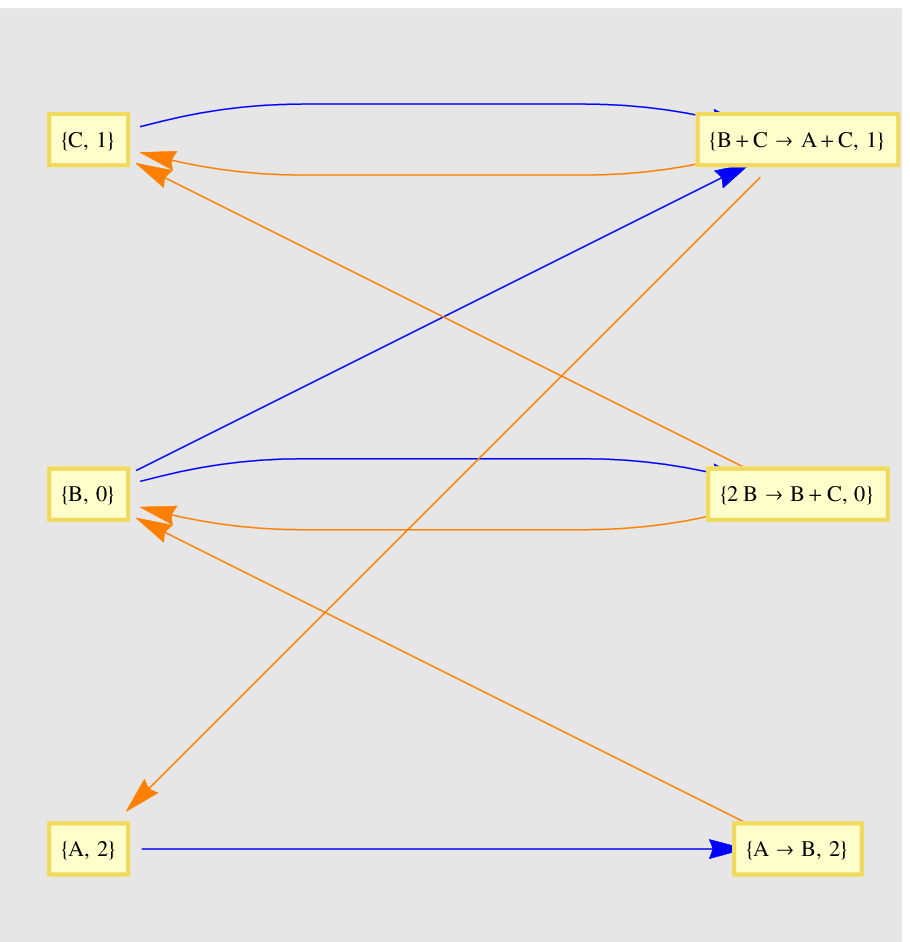}\quad\includegraphics[width=0.26\paperwidth]{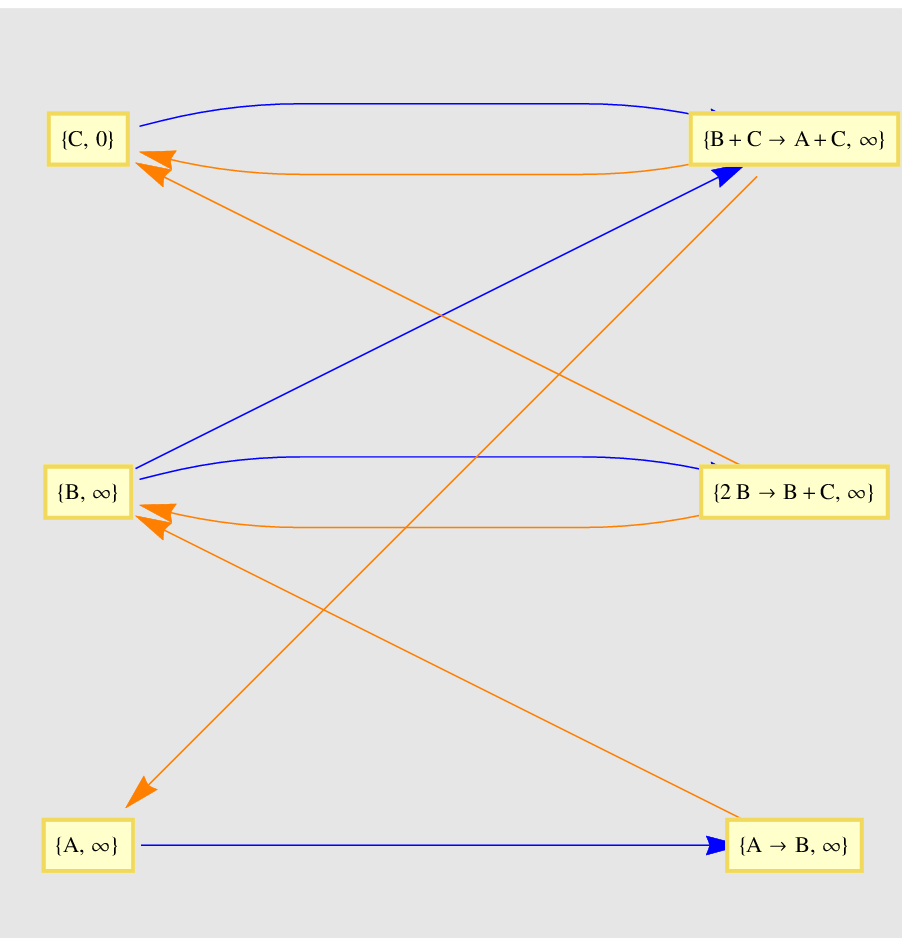}
\caption{The Volpert graphs of the Robertson reaction \eqref{eq:fhjrob} with B and C as initial species, respectively}
\label{fig:RobVgraph}
\end{center}
\end{figure}

Suppose one takes A as the only initial species, then A and the reaction step \ce{A -> B}
gets zero index, B and the reaction steps \ce{2B -> B + C} receives 1, finally C and the reaction step \ce{B + C -> A + C} is assigned 2. Upon selecting B one gets a similar result. However, if one chooses C as the single initial species then all the other species and all the reaction steps will have an infinite index.

One may have the objection that the Robertson mechanism is not detailed balanced (as it is not even reversible).
In some circumstances it may be required that only detailed balanced reactions be taken into consideration, however,
as approximate models one often uses reactions not obeying this principle.
Our view is presented in \cite{nagykovacstoth, nagytoth} in a detailed way.
\end{Exa}

We have done all the calculations of  the characteristic quantities of reactions
using the package \ReactionKinetics\ developed in \textit{Mathematica} and shown also at MaCKiE 2011 \citep{nagypapptoth}
and described in detail in
\cite{springer}.
Figures have also been drawn by the package.

\bibliography{TothNagyZselyCES}%
\bibliographystyle{model5-names}%
\end{document}